\documentclass[twocolumn,american,english,aps,pra,superscriptaddress,nofootinbib]{revtex4-2}
\usepackage[LGR,T1]{fontenc}
\usepackage{textcomp}
\usepackage[utf8]{inputenc}
\usepackage{color}
\usepackage{babel}
\usepackage{amsmath}
\usepackage{amssymb}
\usepackage{graphicx}
\usepackage[bookmarks=false,
 breaklinks=false,pdfborder={0 0 1},colorlinks=true]
 {hyperref}
\hypersetup{
 linkcolor=blue,citecolor=blue,urlcolor=blue}

\makeatletter

\usepackage{color}
\usepackage{amsfonts}\usepackage{bm}\usepackage{graphicx}
\usepackage{mathrsfs}

\makeatother

\begin{document}
\title{Effective Geometry and Position-Dependent Mass in Dual-$q$ Quantum Mechanics}
\author{A. Boumali}
\email{boumali.abdelmalek@gmail.com}

\affiliation{Laboratoire de Physique Appliquée et Théorique, LPAT, Université Larbi-Tébessi,
Tébessa, Algeria}
\author{A. Makhlouf}
\email{abdenacer.makhlouf@uha.fr}

\affiliation{Université de Haute Alsace, IRIMAS-Département de Mathématiques, 68093
Mulhouse, France}
\date{\today}
\selectlanguage{american}%
\begin{abstract}
This work investigates the deformed-derivative formalism introduced by Borges,
with emphasis on the relation between the linear operator $D_{(q)}$ and its
nonlinear dual counterpart $D^{(q)}$. Directly inserting the dual derivative
into the kinetic term leads to a nonlinear Schr\"odinger equation and obscures
the usual interpretation of superposition and probability. We show that this
nonlinearity can be removed by a simultaneous transformation of the coordinate
and of the wave function. The transformed problem is an ordinary linear
Schr\"odinger equation in a deformed coordinate, and its representation in the
physical coordinate is equivalent to a Hermitian position-dependent-mass (PDM)
Hamiltonian. In this formulation, the deformation parameter $q$ determines both
the effective mass profile and the associated metric. The formalism is applied
to the free particle, the infinite square well, the rectangular barrier, and the
harmonic oscillator in the weak-deformation regime. Comparison with the
nonadditive-translation approach of Costa Filho \emph{et al.} shows that the
Borges dual-$q$ framework provides an alternative route to the same effective
geometric structure. For $q<1$, the effective confinement length is reduced,
which raises the bound-state spectrum and enhances tunneling; for $q>1$, the
effective length is increased, which lowers the spectrum and suppresses
tunneling relative to the undeformed limit $q=1$.
\end{abstract}
\maketitle
\selectlanguage{english}%

\section{Introduction}
In recent decades, quantum groups and quantum algebras have become central
structures in mathematical physics. Their development was strongly influenced by
the introduction of $q$-deformed oscillators and by the realization that
$q$-deformed algebras provide nontrivial deformations of universal enveloping
algebras associated with Lie algebras. In the limit $q\to1$, these structures
recover their classical counterparts, while for $q\neq1$ they generate modified
spectra, deformed commutation relations, and altered phase-space geometries. The
realization of $SU_q(2)$ through a $q$-analogue of the harmonic oscillator was a
particularly important step, motivating extensive studies of representations,
$q$-oscillators, and quantum-algebraic models
\cite{Biedenharn1989,Woronowicz1987,ChariPressley1995,KlimykSchmudgen1997,Macfarlane1989,Kulish1990}.

The use of $q$-deformed structures has since expanded to many areas of
theoretical and mathematical physics. Examples include $q$-deformed harmonic and
Morse oscillators, deformed Jaynes--Cummings models, modified oscillator
algebras, $q$-deformed supersymmetric quantum mechanics, deformed ideal-gas
models, Tamm--Dancoff oscillators, and relativistic or molecular oscillator
systems
\cite{AlginSenay2016,AtakishiyevSuslov1990,ArikCoon1976,AtakishiyevSuslov1991,Chargui2025qDKP1p2,Chung2013Tamm,Chung2015Fermi,Chung2015Tamm,ChungHounkonnouArjika2014,Dong1995,Drinfeld1985,Gavrilik2013,Jimbo1985,Klimyk2005,Korichi2022FracDirac,Mancini1994}.
Deformed calculi and generalized algebras are also useful in the description of
complex systems, nonlocal effects, anomalous transport, and effective
geometries
\cite{Serdouk2020Solutions,Serdouk2023Fractional}. A prominent example is the
nonextensive statistical mechanics introduced by Tsallis, in which the entropic
index $q$ measures deviations from the standard Boltzmann--Gibbs framework
\cite{Tsallis1988,Boumali2013GrapheneDO,Boumali2017qDirac,Boumali2018qMorse,Boumali2020SuperstatDO,Boumali2023qDKP,Cai2007,Liu2010,Korichi2022FracDirac}.

Within this broader context, Borges proposed a $q$-calculus that contains a
linear derivative $D_{(q)}$ and an intrinsically nonlinear dual derivative
$D^{(q)}$ \cite{Borges2004}. The operator $D_{(q)}$ has been used in
applications related to generalized statistical mechanics and information
theory. Its dual, however, is less straightforward in quantum mechanics because
it acts nonlinearly on the function to which it is applied. A direct
substitution of $D^{(q)}$ into the kinetic term of a Schr\"odinger equation
therefore produces a nonlinear wave equation and may compromise the usual
linear superposition principle.

A closely related geometric idea appears in position-dependent-mass (PDM)
quantum mechanics. In particular, Costa Filho \emph{et al.} introduced a
nonadditive displacement operator that leads to a deformed momentum operator and
to a Schr\"odinger-like equation interpretable as the dynamics of a particle
with spatially varying effective mass \cite{CostaFilho2011}. More generally,
PDM Hamiltonians require a careful treatment of Hermitian operator ordering and
of the scalar product when the effective mass depends on position
\cite{vonRoos1983,LevyLeblond1995}. In the nonadditive-translation approach,
the deformation parameter controls both the effective confinement length and
the tunneling probability. This connection suggests that the Borges dual
$q$-derivative may also admit a linear and geometrically transparent
interpretation if the correct variables are used.

The purpose of this paper is to construct that interpretation explicitly. We
show that a nonlinear field transformation, combined with a deformed coordinate
map, converts the dual-$q$ Schr\"odinger problem into a standard linear
Schr\"odinger equation in the canonical coordinate. When transformed back to the
physical coordinate, the model becomes a Hermitian PDM Hamiltonian with an
effective mass and metric fixed by the parameter $q$. The resulting framework
is then applied to four representative systems: the free particle, the infinite
square well, the rectangular barrier, and the harmonic oscillator in the
weak-deformation regime. These examples clarify how $q$ acts as a geometric
control parameter that stretches or compresses the effective coordinate and,
consequently, shifts bound-state spectra and modifies tunneling probabilities.

\section{Borges $q$--calculus and its dual derivative}

We briefly recall the $q$--derivative formalism introduced by Borges~\cite{Borges2004}.
For a sufficiently smooth function $f(x)$, the \emph{linear} $q$--derivative
is defined as
\begin{equation}
D_{(q)}f(x)=\big[1+(1-q)x\big]\frac{df}{dx}.\label{eq:BorgesDq}
\end{equation}
\noindent\textit{Dimensional remark.} The combination $1+(1-q)x$ must be dimensionless. Therefore,  one may  regard $x$ as a dimensionless coordinate (e.g., $x\equiv X/\ell_{0}$ for a fixed length scale $\ell_{0}$). Equivalently, one can write $1+(1-q)x\equiv 1+\gamma X$ with $\gamma=(1-q)/\ell_{0}$ having dimensions of inverse length. In the sequel, we keep the compact notation $1+(1-q)x$ with the understanding that $x$ is scaled accordingly.

This operator is linear with respect to $f(x)$ and can be interpreted
as the generator of a nonadditive translation in coordinate $x$.
It reduces to the ordinary derivative in the limit $q\to1$. The \emph{dual}
$q$--derivative is defined by
\begin{equation}
D^{(q)}f(x)=\frac{1}{1+(1-q)\,f(x)}\,\frac{df(x)}{dx},\label{eq:BorgesDualDq}
\end{equation}
and is manifestly nonlinear in $f$: in general $D^{(q)}(cf)\neq c\,D^{(q)}f$
for a constant $c$. The two derivatives obey a duality relation of
the form
\begin{equation}
D^{(q)}f(x)=\frac{1}{\big[1+(1-q)x\big]\big[1+(1-q)f(x)\big]}\,D_{(q)}f(x),
\end{equation}
which intertwines coordinate and field deformations. If the dual derivative
is inserted directly in the kinetic energy term of the Schr\"odinger
equation, the resulting dynamics is nonlinear in the wave function.
In what follows we construct a linearization map that restores linearity
at the level of the equation of motion.

\section{Linearization map and position-dependent mass}

\subsection{Deformed coordinate}

We introduce a deformed coordinate $\xi$ defined by
\begin{equation}
\xi(x)=\frac{1}{1-q}\ln\!\big[1+(1-q)x\big].\label{eq:xiDef}
\end{equation}
This mapping is invertible for $1+(1-q)x>0$ and satisfies
\begin{equation}
\frac{d}{dx}=\frac{1}{1+(1-q)x}\,\frac{d}{d\xi}.\label{eq:dxdxi}
\end{equation}
The transformation~\eqref{eq:xiDef} is similar in structure to the
nonadditive displacement used in Ref.~\cite{CostaFilho2011}, where the translation group is deformed by a parameter $\gamma$ (with dimensions of inverse length). In the scaled-coordinate notation adopted here, one may identify $\gamma\ell_0=1-q$.

\subsection{Field transformation}

Next, we define a transformed wave function $\chi$ by
\begin{equation}
\chi(x)=\frac{1}{1-q}\ln\!\big[1+(1-q)\psi(x)\big],\label{eq:chiDef}
\end{equation}
where $\psi(x)$ is the physical wave function. The inverse relation
is
\begin{equation}
\psi(x)=\frac{1}{1-q}\left[\exp\!\big((1-q)\chi(x)\big)-1\right],\label{eq:psiFromChi}
\end{equation}
which is well defined for $1+(1-q)\psi(x)>0$.

Note that while the auxiliary field $\chi(\xi)$ satisfies the linear
superposition principle, the physical field $\psi(x)$ does not. That
is, if $\chi_{1}$ and $\chi_{2}$ are solutions, the physical state
corresponding to $(\chi_{1}+\chi_{2})$ is not simply $\psi_{1}+\psi_{2}$,
but is given by the nonlinear composition via Eq.~\eqref{eq:psiFromChi}.
Using the chain rule, we obtain
\begin{equation}
D^{(q)}\psi(x)=\frac{d\chi}{dx}.\label{eq:Dqpsi-dchidx}
\end{equation}
Combining this with Eq.~\eqref{eq:dxdxi}, we find
\begin{equation}
D^{(q)}\psi(x)=\frac{d\chi}{dx}=\frac{1}{1+(1-q)x}\,\frac{d\chi}{d\xi}.\label{eq:Dqpsi-dchidxi}
\end{equation}

\subsection{Effective Hamiltonian and PDM form}

The preceding transformations show how the nonlinear field dependence of
$D^{(q)}$ can be absorbed into the auxiliary wave function $\chi$ and the
coordinate $\xi$. A point that must be fixed carefully is the definition of
the Hamiltonian. If one inserts the dual derivative directly into the formal
operator
\begin{equation}
\widetilde K^{(q)}\psi(x)
=-\frac{\hbar^{2}}{2m}\,\partial_x\!\left[D^{(q)}\psi(x)\right],
\label{eq:formal_Kq}
\end{equation}
one obtains, by using Eq.~\eqref{eq:Dqpsi-dchidxi},
\begin{equation}
\widetilde K^{(q)}\psi(x)=
-\frac{\hbar^{2}}{2m}\,\frac{1}{w(x)}\partial_\xi
\left[\frac{1}{w(x)}\partial_\xi\chi(\xi)\right],
\qquad
w(x)=1+(1-q)x .
\label{eq:formal_Kq_explicit}
\end{equation}
This operator is not equal to the canonical free-particle operator
$-(\hbar^2/2m)\partial_\xi^2$. Therefore, Eq.~\eqref{eq:formal_Kq}
will not be used as the physical Hamiltonian of the present model. It is only
the formal result obtained from a direct differential substitution.

The physical assumption adopted in this work is instead that the deformed
coordinate $\xi$ is the canonical coordinate. The momentum conjugate to
$\xi$ is
\begin{equation}
\hat p_\xi=-i\hbar\frac{\partial}{\partial\xi},
\label{eq:pxi_def}
\end{equation}
and the Hamiltonian acting on the linearized wave function $\chi(\xi)$ is
\begin{equation}
\hat H_\xi=-\frac{\hbar^{2}}{2m}\frac{d^2}{d\xi^2}+V\big(x(\xi)\big).
\label{eq:Hxi_def}
\end{equation}
Thus the stationary equation is
\begin{equation}
-\frac{\hbar^{2}}{2m}\,\frac{d^{2}\chi}{d\xi^{2}}
+V\big(x(\xi)\big)\chi(\xi)=E\chi(\xi).
\label{eq:Schro-xi}
\end{equation}
With this choice the free particle remains a plane wave in the canonical
coordinate, and the deformation enters through the coordinate map and the
corresponding measure.

The same dynamics can be expressed in the physical coordinate $x$ as a
position-dependent-mass (PDM) problem. Since
\begin{equation}
 d\xi=\frac{dx}{w(x)},
\end{equation}
probability conservation requires
\begin{equation}
|\chi(\xi)|^2d\xi=|\phi(x)|^2dx,
\qquad
\chi(\xi(x))=w^{1/2}(x)\phi(x),
\label{eq:chi_phi}
\end{equation}
where $\phi$ is the equivalent linear wave function normalized with the flat
measure $dx$. Using $\partial_\xi=w(x)\partial_x$, the kinetic part of
Eq.~\eqref{eq:Hxi_def} becomes
\begin{equation}
\hat K_x\phi(x)=
-\frac{\hbar^{2}}{2m}\,w^{1/2}(x)\frac{d}{dx}
\left[w(x)\frac{d}{dx}\left(w^{1/2}(x)\phi(x)\right)\right].
\label{eq:PDM_w_form}
\end{equation}
Equivalently, by defining
\begin{equation}
 m_{\mathrm{eff}}(x)=\frac{m}{w^2(x)},
\label{eq:meff_correct}
\end{equation}
we obtain the von Roos Hamiltonian
\begin{equation}
\hat K_x=-\frac{\hbar^{2}}{2}\,
 m_{\mathrm{eff}}^{-1/4}(x)\frac{d}{dx}
\left[m_{\mathrm{eff}}^{-1/2}(x)\frac{d}{dx}
 m_{\mathrm{eff}}^{-1/4}(x)\right].
\label{eq:PDM_Final}
\end{equation}
This corresponds to the Hermitian ordering
$\alpha=\gamma=-1/4$ and $\beta=-1/2$, satisfying
$\alpha+\beta+\gamma=-1$. The ordering is not arbitrary: it follows from
the unitary transformation between the measures $d\xi$ and $dx$.

This result establishes the connection with the nonadditive translation
formalism of Costa Filho \emph{et al.}~\cite{CostaFilho2011}. Their
effective mass has the structure $m_e(x)\propto(1+\gamma x)^{-2}$. In the
present notation the correspondence is
\begin{equation}
\gamma = 1-q,
\end{equation}
so that the two descriptions encode the same deformed geometry, although
they arise from different starting points.

Finally, in the $\xi$ representation the density and current satisfy
\begin{equation}
\partial_t|\chi|^2+\partial_\xi j_\xi=0,
\qquad
j_\xi=\frac{\hbar}{m}\,\Im\left(\chi^*\frac{d\chi}{d\xi}\right).
\end{equation}
In the flat $x$-space representation,
\begin{equation}
\rho_x(x)=|\phi(x)|^2=\frac{|\chi(\xi(x))|^2}{w(x)},
\qquad
J_x(x)=j_\xi(\xi(x)),
\end{equation}
and one obtains $\partial_t\rho_x+\partial_xJ_x=0$. Therefore the
$q$-deformation modifies the probability density through the Jacobian while
preserving unitary probability flow.

\subsection{Standard $q$-derivative: review}

It is instructive to compare the results obtained via the dual derivative
$D^{(q)}$ with those arising from the standard, linear $q$-derivative
$D_{(q)}$. Recall that the standard $q$-derivative is defined by~\cite{Borges2004}:
\begin{equation}
D_{(q)}f(x)=\left[1+(1-q)x\right]\frac{df}{dx}.
\end{equation}
Unlike its dual partner, $D_{(q)}$ is a linear operator. Consequently,
a kinetic energy term constructed from it, $\hat{K}_{(q)}\propto[D_{(q)}]^{2}$,
leads to a linear differential equation without the need for the auxiliary
field transformation $\psi\to\chi$. We may proceed using only the
coordinate transformation $x\to\xi$ defined in Eq.~\eqref{eq:xiDef}.
Substituting the relation $\partial_{\xi}=[1+(1-q)x]\partial_{x}$
into the definition of $D_{(q)}$, we find simply that $D_{(q)}=\partial_{\xi}$.
If we postulate the canonical quantization directly in the $\xi$
space as before, the Schr\"odinger equation becomes:
\begin{equation}
-\frac{\hbar^{2}}{2m}\frac{d^{2}\psi}{d\xi^{2}}=E\psi(\xi).
\end{equation}
This is mathematically identical to Eq.~\eqref{eq:Schro-xi}, but
now for the physical field $\psi$ itself rather than the auxiliary
field $\chi$.
\begin{itemize}
\item For the Infinite Potential Well, the boundary conditions in the $\xi$
domain are identical ($\psi(0)=\psi(\Xi)=0$). Therefore, the energy
eigenvalues $E_{n}^{(q)}$ are exactly the same as those derived in
Eq.~\eqref{eq:En-q}. However, the physical wave functions differ
significantly. In the standard $q$-derivative case, the wave function
is:
\begin{equation}
\psi_{n}^{(std)}(x)\propto\sin\left(\frac{n\pi\xi(x)}{\Xi(q)}\right)=\sin\left(\frac{n\pi\ln[1+(1-q)x]}{(1-q)\Xi(q)}\right).
\end{equation}
This stands in contrast to the dual case, where $\psi_{n}(x)$ involves
the exponential nonlinearity $\exp(\sin\xi)-1$.
\item For the Rectangular Barrier, the transmission coefficient depends
only on the effective width of the barrier in the canonical coordinate.
Since the mapping $x\to\xi$ determines the effective width $A(q)$
identically in both cases, the transmission probability $T(E;q)$
is the same as Eq.~\eqref{eq:Tunneling-q}.
\end{itemize}
In summary, the geometric effects of the deformation---specifically
the shifting of energy levels and the modulation of tunneling probabilities---are
robust features arising from the deformation of the coordinate space.
The specific form of the wave function, however, depends on whether
the underlying derivative operator requires field linearization (dual
case) or is intrinsically linear (standard case).

\section{Applications}

\subsection{Free particle in the dual-$q$ framework}

Before turning to bounded and scattering states, we briefly discuss the free
particle. For $V(x)=0$, Eq.~\eqref{eq:Schro-xi} becomes
\begin{equation}
-\frac{\hbar^{2}}{2m}\frac{d^{2}\chi}{d\xi^{2}}=E\chi,
\end{equation}
with generalized eigenfunctions
\begin{equation}
\chi_k(\xi)=\frac{1}{\sqrt{2\pi}}e^{ik\xi},
\qquad
E=\frac{\hbar^2k^2}{2m}.
\label{eq:free-xi}
\end{equation}
The domain condition $w(x)=1+(1-q)x>0$ restricts the allowed range of
$x$, but the logarithmic map sends the singular endpoint to infinity in
$\xi$. For $q<1$, the endpoint $x=-1/(1-q)$ corresponds to
$\xi\to-\infty$; for $q>1$, the endpoint $x=1/(q-1)$ corresponds to
$\xi\to+\infty$. Hence the canonical free-particle problem remains defined
on the full real $\xi$ line and no additional finite-wall boundary condition
is imposed at the singular point. In the physical coordinate, the wave
profile is distorted by $\xi(x)$ and by the nonlinear inverse field map,
\begin{equation}
\psi_k(x)=\frac{1}{1-q}\left[\exp\!\left((1-q)\chi_k(\xi(x))\right)-1\right],
\end{equation}
but the dispersion relation remains the standard one in the canonical
coordinate. This is the appropriate free-particle limit of the present
canonical $\xi$-space formulation.

\subsection{Infinite potential well}

We now consider a particle confined in the interval $0<x<L$, with
\begin{equation}
V(x)=\begin{cases}
0, & 0<x<L,\\
\infty, & \text{otherwise},
\end{cases}
\end{equation}
and Dirichlet boundary conditions $\psi(0)=\psi(L)=0$. In the deformed
coordinate $\xi$, defined by Eq.~\eqref{eq:xiDef}, the interval
$[0,L]$ is mapped to $[0,\Xi(q)]$, where
\begin{equation}
\Xi(q)=\xi(L)-\xi(0)=\frac{1}{1-q}\ln\!\big[1+(1-q)L\big].\label{eq:Xi-q}
\end{equation}
Inside the well, the time-independent equation in $\xi$ reads
\begin{equation}
-\frac{\hbar^{2}}{2m}\,\frac{d^{2}\chi}{d\xi^{2}}=E\,\chi,
\end{equation}
with boundary conditions $\chi(0)=\chi(\Xi)=0$. The normalized eigenfunctions
are
\begin{equation}
\chi_{n}(\xi)=\sqrt{\frac{2}{\Xi(q)}}\,\sin\!\left(\frac{n\pi\xi}{\Xi(q)}\right),\qquad n=1,2,\dots
\end{equation}
and the quantized energy levels are
\begin{equation}
E_{n}^{(q)}=\frac{\hbar^{2}\pi^{2}n^{2}}{2m\,\Xi^{2}(q)}=\frac{\hbar^{2}\pi^{2}n^{2}}{2m}\left[\frac{1-q}{\ln\big(1+(1-q)L\big)}\right]^{2}.\label{eq:En-q}
\end{equation}
For the case $q>1$, the well width $L$ is physically constrained
by the coordinate singularity. Existence of the well requires $L<1/(q-1)$
to ensure the argument of the logarithm remains positive. The $q$--dependence
of $\Xi(q)$ and $E_{n}^{(q)}$ mirrors the $\gamma$--dependence
of the effective length obtained from the translation operator in
Ref.~\cite{CostaFilho2011}. In that work, the energy levels scale
as $E_{n}\propto\gamma^{2}/\ln^{2}(1+\gamma L)$. Comparing this with
Eq.~\eqref{eq:En-q}, we see the exact isomorphism with $\gamma\leftrightarrow1-q$.
For $q<1$ (corresponding to $\gamma>0$), one has $\Xi(q)<L$, so
the spectrum is shifted upward relative to the undeformed case $q=1$
(``blue shift''). For $q>1$, the effective width is larger than
$L$ and the levels become more closely spaced (``red shift'').
The physical wave functions $\psi_{n}(x)$ are obtained from Eq.~\eqref{eq:psiFromChi}:
\begin{equation}
\psi_{n}(x)=\frac{1}{1-q}\left\{ \exp\!\Big[(1-q)\chi_{n}\big(\xi(x)\big)\Big]-1\right\} ,
\end{equation}
with $\xi(x)$ given by Eq.~\eqref{eq:xiDef}. In general, the profiles
$\psi_{n}(x)$ are asymmetric in the physical coordinate, reflecting
the deformation of the underlying geometry, similarly to the result
in Ref.~\cite{CostaFilho2011}.

\begin{figure}
\begin{centering}
\includegraphics[scale=0.4]{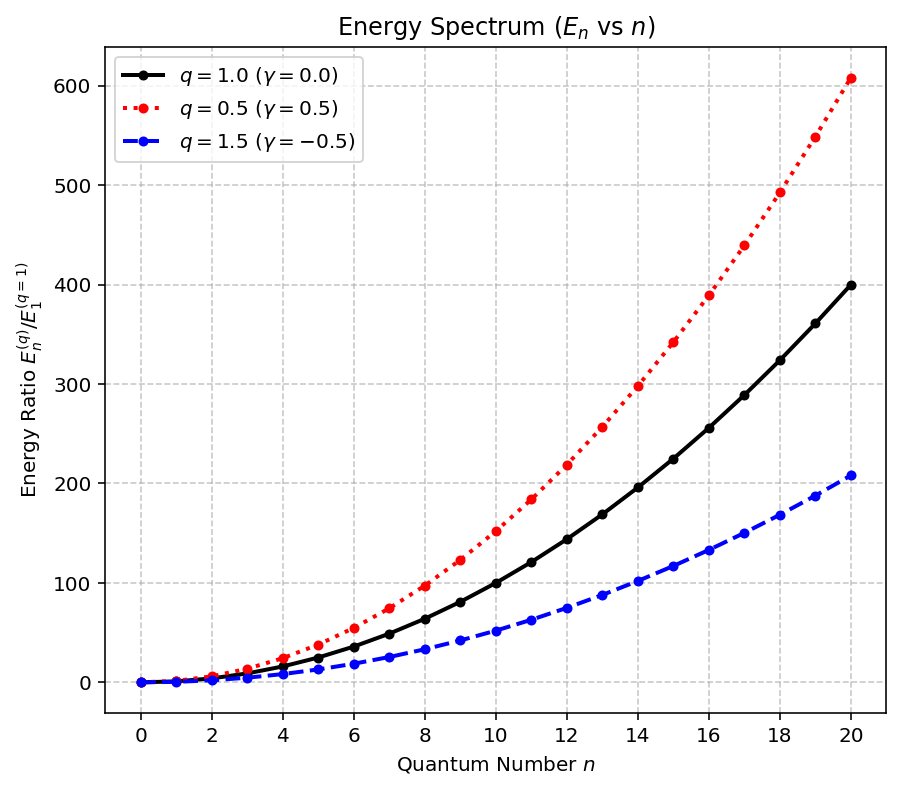}
\par\end{centering}
\caption{Energy Spectrum of the Dual-$q$ Infinite Well. Energy levels $E_{n}^{(q)}$
(normalized to $E_{1}^{(q=1)}$) versus quantum number $n$ for well
width $L=1.0$. The black solid line represents the standard case
($q=1.0$). The red dotted line ($q=0.5$) shows a significant energy
increase (blue shift) due to effective spatial compression ($\Xi(q)<L$).
The blue dashed line ($q=1.5$) shows energy suppression (red shift)
corresponding to spatial dilation ($\Xi(q)>L$)}\label{fig:Energy}
\end{figure}
\begin{figure}
\begin{centering}
\includegraphics[scale=0.3]{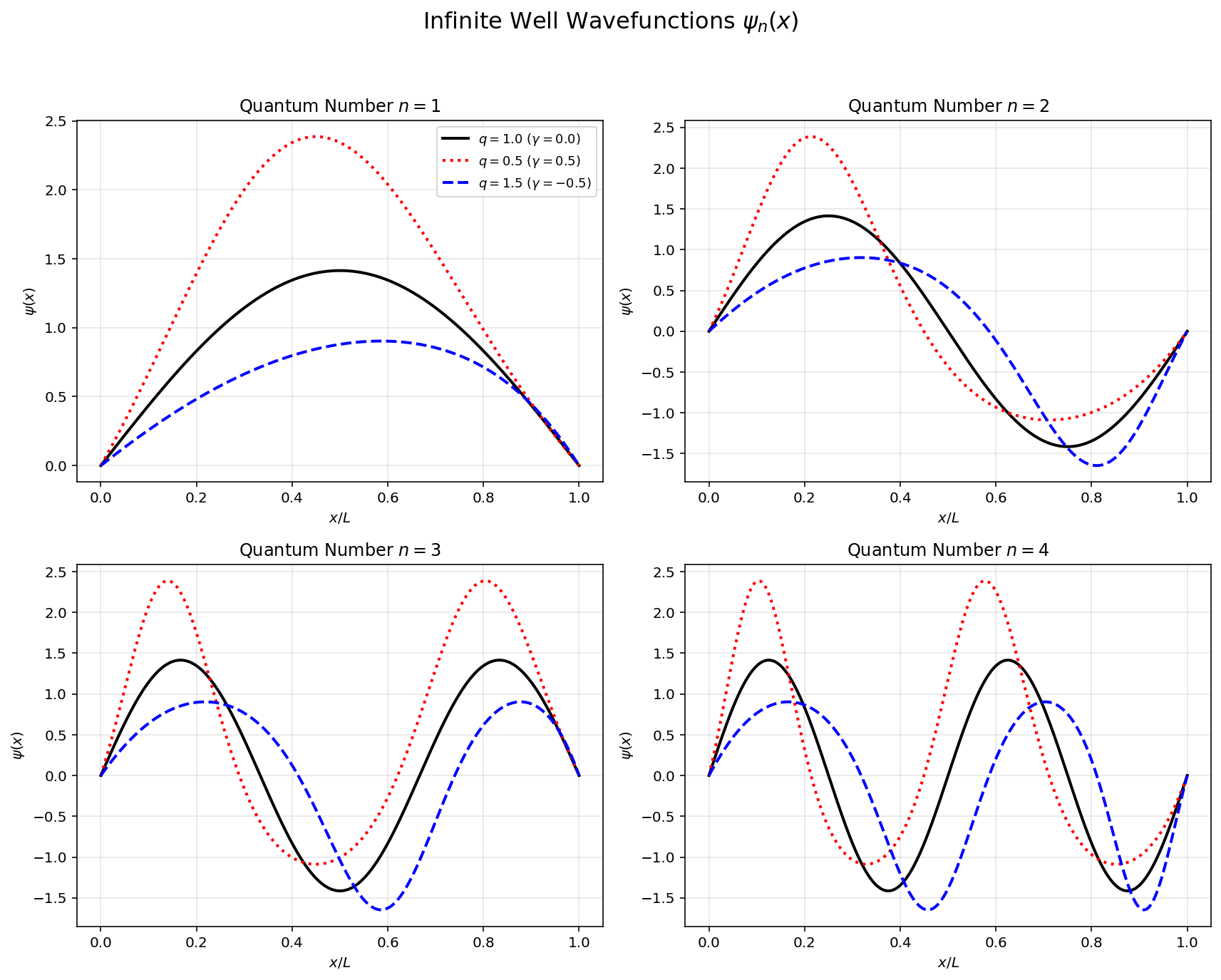}
\par\end{centering}
\caption{Wave function Geometry and Spatial Deformation. Physical wave functions
$\psi_{n}(x)$ for the ground state ($n=1$) and excited states. For
$q=1.0$ (black), the functions are standard symmetric standing waves.
For $q=0.5$ (red), the wave functions are skewed to the left, reflecting
the compression of the metric. For $q=1.5$ (blue), they are skewed
to the right, reflecting dilation. This asymmetry is directly responsible
for the deviation of the average position $\langle x\rangle$ from
$L/2$.}\label{fig:Wave function}
\end{figure}

The dependence of the energy spectrum on the deformation parameter
is illustrated in Fig.~\ref{fig:Energy}.
For $q<1$ (equivalent to $\gamma>0$ in Ref.~\cite{CostaFilho2011}),
the effective width $\Xi(q)$ is smaller than $L$. This compression
of the coordinate space leads to a ``blue shift'' in the energy levels,
which grow more rapidly with $n$ compared to the standard case ($q=1$).
Conversely, for $q>1$ (equivalent to $\gamma<0$), the effective
width is larger than $L$, resulting in a ``red shift'' and a compressed
energy spectrum.

The physical wave functions $\psi_{n}(x)$ are obtained from Eq.~\eqref{eq:psiFromChi}.
The geometric deformation also manifests in the spatial distribution
of the probability density, as shown in Fig.~\ref{fig:Wave function}.
For $q\neq1$, the wave functions lose their symmetry around $L/2$.
For $q=0.5$ (red), the density shifts toward the region where the
effective mass is larger (small $x$), while for $q=1.5$ (blue),
it is pushed toward the region of smaller effective mass.

\subsubsection{Thermal properties}

\subsubsection{Partition function for the dual-$q$ infinite square well: Mellin transform, zeta function and residue evaluation}

For the infinite barrier (infinite square well) in the dual-$q$ framework, the energy spectrum reads
\begin{equation}
E_n^{(q)}=\varepsilon_q\,n^2,\qquad n=1,2,\dots,
\label{eq:En_dualq_well}
\end{equation}
with
\begin{equation}
\varepsilon_q=\frac{\hbar^2\pi^2}{2m\,\Xi(q)^2},
\qquad
\Xi(q)=\frac{1}{1-q}\ln\!\big(1+(1-q)L\big).
\label{eq:epsq_Xiq}
\end{equation}
The canonical partition function is therefore
\begin{equation}
Z_q(\beta)=\sum_{n=1}^{\infty}\exp\!\big(-\beta \varepsilon_q n^2\big),
\qquad
a\equiv \beta\varepsilon_q>0.
\label{eq:Zq_def}
\end{equation}

\paragraph{Mellin representation and zeta factorization}

We use the Mellin transform identity (inverse Mellin representation)
\begin{equation}
e^{-x}=\frac{1}{2\pi i}\int_{c-i\infty}^{c+i\infty} x^{-s}\Gamma(s)\,ds,
\qquad c>0.
\label{eq:mellin_exp}
\end{equation}
Setting $x=a n^2$ and summing over $n$ yields, for $\Re(s)$ large enough,
\begin{align}
Z_q(\beta)
&=\sum_{n=1}^{\infty}\frac{1}{2\pi i}\int_C (a n^2)^{-s}\Gamma(s)\,ds
\nonumber\\
&=\frac{1}{2\pi i}\int_C a^{-s}\Gamma(s)\sum_{n=1}^{\infty}n^{-2s}\,ds
=\frac{1}{2\pi i}\int_C a^{-s}\Gamma(s)\zeta(2s)\,ds.
\label{eq:Z_mellin_zeta}
\end{align}
Hence
\begin{equation}
Z_q(\beta)=\frac{1}{2\pi i}\int_{C} (\beta\varepsilon_q)^{-s}\,\Gamma(s)\,\zeta(2s)\,ds.
\label{eq:Zq_mellin_final}
\end{equation}

\paragraph{Poles of $\Gamma(s)$ and residues of the integrand}

Define the integrand
\begin{equation}
f(s)=a^{-s}\Gamma(s)\zeta(2s).
\label{eq:integrand_def}
\end{equation}
Its singular structure is:

\begin{itemize}
\item $\Gamma(s)$ has simple poles at $s=0,-1,-2,\dots$ with residues
\begin{equation}
\operatorname*{Res}_{s=-n}\Gamma(s)=\frac{(-1)^n}{n!},
\qquad n=0,1,2,\dots.
\label{eq:gamma_res}
\end{equation}

\item $\zeta(2s)$ has a simple pole at $2s=1$, i.e. $s=\tfrac12$, with
\begin{equation}
\zeta(2s)\sim \frac{1}{2s-1}+\cdots
\quad\Rightarrow\quad
\operatorname*{Res}_{s=1/2}\zeta(2s)=\frac12,
\label{eq:zeta_res_half}
\end{equation}
and has trivial zeros at $2s=-2,-4,\dots$, i.e. $s=-1,-2,\dots$:
\begin{equation}
\zeta(-2n)=0,\qquad n=1,2,\dots.
\label{eq:zeta_trivial_zeros}
\end{equation}
\end{itemize}

\paragraph{Residue at $s=\tfrac12$.}
Using $\Gamma(\tfrac12)=\sqrt{\pi}$ and \eqref{eq:zeta_res_half},
\begin{equation}
\operatorname*{Res}_{s=1/2} f(s)
=a^{-1/2}\Gamma(1/2)\operatorname*{Res}_{s=1/2}\zeta(2s)
=\frac{\sqrt{\pi}}{2}\,a^{-1/2}.
\label{eq:res_half}
\end{equation}

\paragraph{Residues at all $\Gamma$-poles $s=-n$ ($n=0,1,2,\dots$).}
Near $s=-n$,
\begin{equation}
\Gamma(s)\sim \frac{(-1)^n}{n!}\frac{1}{s+n}+\cdots,
\end{equation}
so
\begin{equation}
\operatorname*{Res}_{s=-n} f(s)=a^{n}\frac{(-1)^n}{n!}\,\zeta(-2n).
\label{eq:res_minus_n_general}
\end{equation}
Now, \eqref{eq:zeta_trivial_zeros} implies that \emph{all} residues at negative integers vanish:
\begin{equation}
\operatorname*{Res}_{s=-n} f(s)=0,\qquad n=1,2,3,\dots
\label{eq:res_neg_cancel}
\end{equation}
because the $\Gamma$-poles are canceled by the trivial zeros of $\zeta(2s)$.
The only surviving $\Gamma$-pole contribution is at $s=0$:
\begin{equation}
\operatorname*{Res}_{s=0} f(s)=a^{0}\operatorname*{Res}_{s=0}\Gamma(s)\,\zeta(0)=1\cdot \zeta(0)=-\frac12.
\label{eq:res_zero}
\end{equation}

\paragraph{High-temperature (small-$\beta$) expansion from residues}

Closing the contour to the left (standard Mellin inversion argument) gives the leading contribution as the sum of the residues
at $s=\tfrac12$ and $s=0$, while the remaining part is exponentially small for $a\to 0^+$:
\begin{equation}
Z_q(\beta)=\frac{\sqrt{\pi}}{2}\,(\beta\varepsilon_q)^{-1/2}-\frac12+\mathcal{R}(\beta),
\qquad
\mathcal{R}(\beta)=\mathcal{O}\!\left(e^{-\pi^2/(\beta\varepsilon_q)}\right).
\label{eq:Zq_asympt}
\end{equation}
Using \eqref{eq:epsq_Xiq},
\begin{equation}
(\beta\varepsilon_q)^{-1/2}
=\Xi(q)\sqrt{\frac{2m}{\beta\hbar^2\pi^2}},
\end{equation}
and thus
\begin{equation}
Z_q(\beta)
=\Xi(q)\sqrt{\frac{m}{2\pi\hbar^2\beta}}-\frac12+\mathcal{R}(\beta).
\label{eq:Zq_asympt_Xi}
\end{equation}

\paragraph{Exact representation (Jacobi theta / Poisson resummation)}

The sum \eqref{eq:Zq_def} is related to the Jacobi theta function:
\begin{equation}
Z_q(\beta)=\sum_{n=1}^{\infty}e^{-a n^2}
=\frac{1}{2}\Big(\vartheta_3(0,e^{-a})-1\Big).
\label{eq:Z_theta3}
\end{equation}
Using the Jacobi inversion formula (equivalently Poisson resummation),
one obtains the exact identity
\begin{equation}
\sum_{n=1}^{\infty}e^{-a n^2}
=
\frac{1}{2}\left(\sqrt{\frac{\pi}{a}}-1\right)
+\sqrt{\frac{\pi}{a}}\sum_{n=1}^{\infty}\exp\!\left(-\frac{\pi^2 n^2}{a}\right),
\qquad a>0.
\label{eq:Z_exact_poisson}
\end{equation}
The first two terms in \eqref{eq:Z_exact_poisson} reproduce precisely the residue result
\eqref{eq:Zq_asympt}, and the remaining series gives the exponentially small remainder $\mathcal{R}(\beta)$.

\subsubsection{Thermodynamic functions and specific heat}

In the canonical ensemble, the fundamental thermodynamic function is the partition function
\begin{equation}
Z_q(\beta)=\sum_{n=1}^{\infty}e^{-\beta E_n^{(q)}}
=\sum_{n=1}^{\infty}\exp\!\big(-\beta\,\varepsilon_q n^2\big),
\qquad \beta=\frac{1}{k_B T}.
\label{eq:Zq_canonical}
\end{equation}
Once $Z_q$ is known, the Helmholtz free energy, internal energy, and entropy follow as
\begin{equation}
F_q(\beta)=-\frac{1}{\beta}\ln Z_q(\beta),\qquad
U_q(\beta)=-\frac{\partial}{\partial\beta}\ln Z_q(\beta),\qquad
S_q=k_B\big(\ln Z_q+\beta U_q\big).
\label{eq:FUS_from_Z}
\end{equation}
For the dual-$q$ infinite well with $E_n^{(q)}=\varepsilon_q n^2$, the internal energy may be written explicitly as
\begin{equation}
U_q(\beta)=
\frac{\sum_{n=1}^{\infty}\varepsilon_q n^2\,e^{-\beta\varepsilon_q n^2}}
     {\sum_{n=1}^{\infty}e^{-\beta\varepsilon_q n^2}}.
\label{eq:Uq_ratio}
\end{equation}
The specific heat at constant volume is obtained from
\begin{equation}
C_{V,q}(T)=\left(\frac{\partial U_q}{\partial T}\right)_V
= k_B\beta^2\Big(\langle E^2\rangle-\langle E\rangle^2\Big),
\label{eq:Cv_fluctuation}
\end{equation}
where $\langle\cdot\rangle$ denotes canonical averages.

\begin{figure}
\begin{centering}
\includegraphics[width=\linewidth]{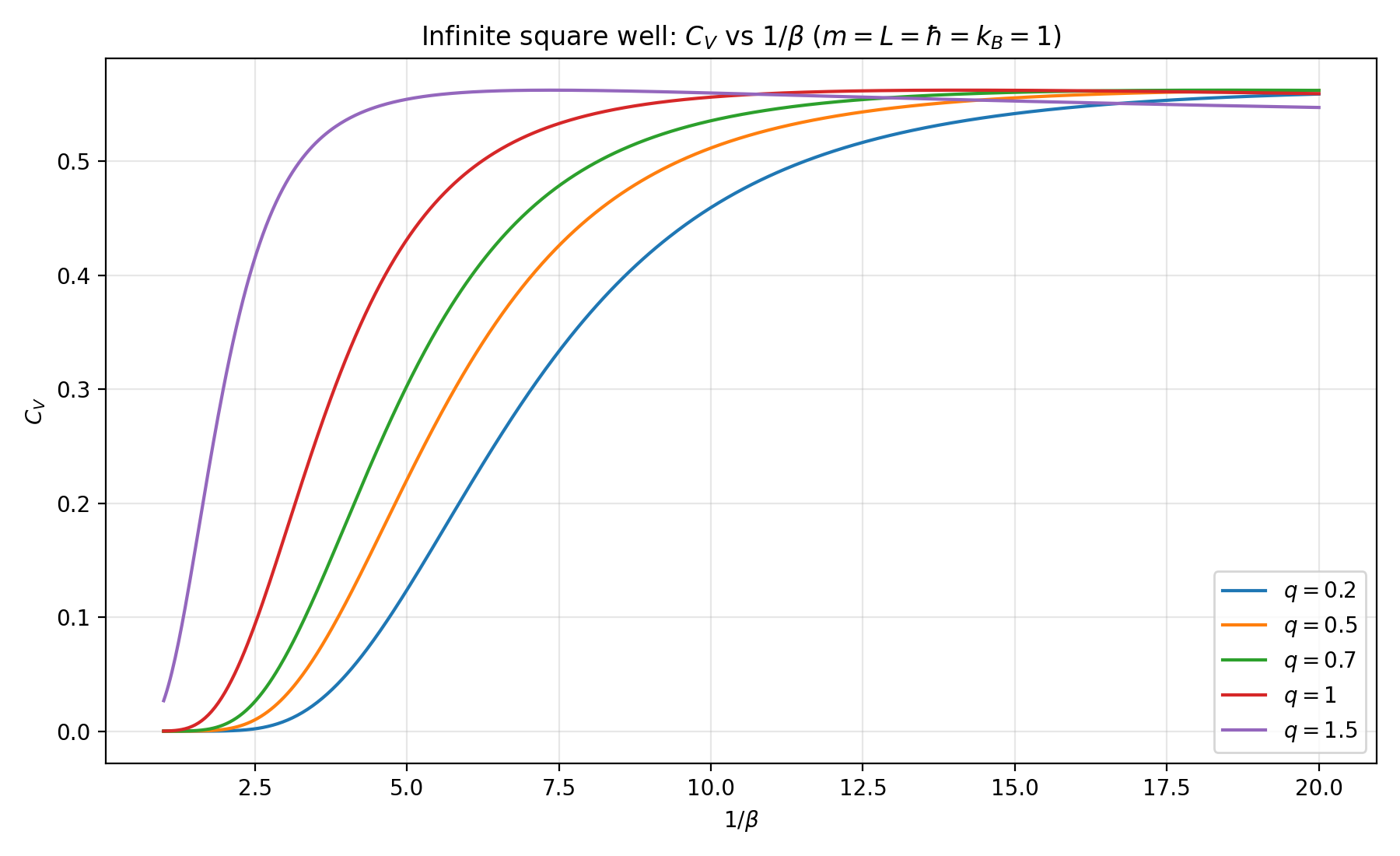}
\par\end{centering}
\caption{Specific heat $C_{V,q}$ versus $1/\beta$ for the dual-$q$ infinite square well with
$q=0.2,\,0.5,\,0.7,\,1.0,\,1.5$ in units $m=L=\hbar=k_B=1$.
All curves approach the classical plateau $C_V\to \tfrac12 k_B$ at large $1/\beta$ (high temperature).
For $q<1$, the effective energy scale $\varepsilon_q$ increases and the rise of $C_V$ is shifted toward larger $1/\beta$,
while for $q>1$ the smaller level spacing yields an earlier onset and faster saturation.}
\label{fig:Cv_invbeta_dualq}
\end{figure}

\paragraph{Comments and classical limit.}
The specific heat $C_{V,q}$ for the dual-$q$ infinite square well, as shown in Fig.~\ref{fig:Cv_invbeta_dualq}, exhibits characteristic behavior influenced by the deformation parameter $q$. At low temperatures (small $1/\beta$), $C_{V,q}$ starts from zero because the system is predominantly in the ground state, with minimal thermal excitation. As temperature increases, $C_{V,q}$ rises and approaches a $q$-independent plateau of $\frac{1}{2}k_B$ at high temperatures, consistent with the classical equipartition theorem for a one-dimensional system with a single quadratic degree of freedom (kinetic energy).

The deformation parameter $q$ acts as a modulator of the effective energy scale $\varepsilon_q \propto [\Xi(q)]^{-2}$. For $q < 1$, the effective well width $\Xi(q) < L$, leading to larger level spacings ($\varepsilon_q > \varepsilon_{q=1}$). This results in delayed thermal excitation, shifting the rise of $C_{V,q}$ to higher temperatures (larger $1/\beta$). Conversely, for $q > 1$, $\Xi(q) > L$, reducing the level spacing ($\varepsilon_q < \varepsilon_{q=1}$), which causes an earlier onset of the specific heat increase and faster saturation to the classical limit.

This $q$-dependence highlights how the deformation parameter effectively rescales the confinement geometry, altering the thermal response. In the high-temperature limit, the system behaves classically, independent of $q$, as quantum effects become negligible, and the equipartition theorem dominates. This behavior underscores the interplay between quantum deformation and thermodynamic properties, providing insights into how geometric modifications influence heat capacity in confined systems.

The plateau corresponds to the classical (equipartition) limit. In the classical continuum approximation,
the translational Hamiltonian is quadratic in the momentum, $H=p^2/(2m)$, and the canonical energy satisfies
$U\simeq \tfrac{1}{2}k_B T$ in one dimension; therefore
\begin{equation}
C_V \;\xrightarrow[T\to\infty]{}\; \frac{1}{2}k_B,
\label{eq:classical_Cv_halfkB}
\end{equation}
in agreement with the asymptotic behavior of all curves in Fig.~\ref{fig:Cv_invbeta_dualq}.
This result is the one-particle, one-dimensional analog of classical equipartition. By contrast, the
Dulong--Petit law applies to crystalline solids in three dimensions, where each atom behaves (approximately)
as a set of three independent harmonic vibrational modes, yielding $C_V\simeq 3k_B$ per atom (or $3Nk_B$ for $N$
atoms) at high temperature. Our infinite-well system has only one quadratic kinetic degree of freedom per particle,
so its classical limit is $\tfrac{1}{2}k_B$ rather than the Dulong--Petit value.

\subsection{Rectangular barrier and tunneling}

We next analyze tunneling through a rectangular potential barrier
of height $V_{0}$ and width $a$, located in the region $0<x<a$:
\begin{equation}
V(x)=\begin{cases}
V_{0}, & 0<x<a,\\
0, & \text{otherwise}.
\end{cases}
\end{equation}
In the deformed coordinate, the barrier occupies the interval $[0,A(q)]$,
where
\begin{equation}
A(q)=\xi(a)-\xi(0)=\frac{1}{1-q}\ln\!\big[1+(1-q)a\big].\label{eq:A-q}
\end{equation}
The stationary Schr\"odinger equation in $\xi$ is identical to the
textbook problem of a rectangular barrier of width $A(q)$ and height
$V_{0}$. To determine explicitly the reflection coefficient $R$
without invoking the unitarity relation $R+T=1$, we employ the transfer
matrix method directly in the deformed coordinate $\xi$. As established,
the physical barrier maps to a standard rectangular barrier of height
$V_{0}$ and effective width $A(q)$. We consider a particle with
energy $E<V_{0}$ incident from the left ($\xi<0$). The wave functions
in the three relevant regions of the deformed space are given by:
\begin{equation}
\chi(\xi)=\begin{cases}
e^{ik\xi}+re^{-ik\xi}, & \xi<0\quad(\text{Region I})\\
C\cosh(\kappa\xi)+D\sinh(\kappa\xi), & 0<\xi<A(q)\quad(\text{Region II})\\
te^{ik\xi}, & \xi>A(q)\quad(\text{Region III})
\end{cases}
\end{equation}
where the wave numbers are defined as $k=\sqrt{2mE}/\hbar$ and $\kappa=\sqrt{2m(V_{0}-E)}/\hbar$. By imposing the standard continuity conditions for the wave function
$\chi(\xi)$ and its derivative $d\chi/d\xi$ at the boundaries $\xi=0$
and $\xi=A(q)$, we obtain the following system of equations:
\begin{equation}
\begin{aligned}1+r & =C,\\
ik(1-r) & =\kappa D,\\
C\cosh(\kappa A)+D\sinh(\kappa A) & =te^{ikA},\\
\kappa C\sinh(\kappa A)+\kappa D\cosh(\kappa A) & =ikte^{ikA},
\end{aligned}
\label{eq:BarrierSystem}
\end{equation}
where $A\equiv A(q)$ for notational brevity. Solving for the reflection
amplitude $r$ by eliminating the transmission amplitude $t$ and
the coefficients $C$ and $D$ yields:
\begin{equation}
r=\frac{(k^{2}+\kappa^{2})\sinh(\kappa A)}{(k^{2}-\kappa^{2})\sinh(\kappa A)+2ik\kappa\cosh(\kappa A)}.
\end{equation}
The reflection coefficient is then calculated as $R=|r|^{2}$. Using
the hyperbolic identity $\cosh^{2}(x)=1+\sinh^{2}(x)$, the modulus
squared simplifies to:
\begin{equation}
R=\frac{(k^{2}+\kappa^{2})^{2}\sinh^{2}(\kappa A)}{(k^{2}+\kappa^{2})^{2}\sinh^{2}(\kappa A)+4k^{2}\kappa^{2}}.
\end{equation}
Substituting the explicit expressions for $k$ and $\kappa$, the
prefactor reduces to:
\begin{equation}
\frac{(k^{2}+\kappa^{2})^{2}}{4k^{2}\kappa^{2}}=\frac{V_{0}^{2}}{4E(V_{0}-E)}.
\end{equation}
Finally, the reflection coefficient in the dual-$q$ formalism is
given by:
\begin{equation}
R(E;q)=\frac{\frac{V_{0}^{2}}{4E(V_{0}-E)}\sinh^{2}(\kappa A(q))}{1+\frac{V_{0}^{2}}{4E(V_{0}-E)}\sinh^{2}(\kappa A(q))}.
\end{equation}
This result allows for a direct verification of probability conservation,
$R(E;q)+T(E;q)=1$. Similarly, we can explicitly derive the transmission
coefficient $T(E;q)$ from the system of equations~\eqref{eq:BarrierSystem}.
To eliminate $C$ and $D$ and solve for the transmission amplitude
$t$, we can rewrite the first two equations as:
\begin{align}
C & =\frac{ik(1+r)+ik(1-r)}{2ik}+\frac{\kappa D}{2ik}.
\end{align}
A more direct approach is to eliminate $r$ first. From the first equation
$C=1+r$, so $r=C-1$. Substituting into the second equation: $ik[1-(C-1)]=\kappa D\implies ik(2-C)=\kappa D$.
This gives $2ik=ikC+\kappa D$. Now we express $C$ and $D$ in terms
of $t$ using the boundary conditions at $\xi=A$.

\begin{widetext}
Multiplying the
third equation by $\kappa$ and adding/subtracting the fourth equation
yields:
\begin{align}
\kappa(C\cosh\kappa A+D\sinh\kappa A)+(\kappa C\sinh\kappa A+\kappa D\cosh\kappa A) & =\kappa te^{ikA}+ikte^{ikA},\nonumber \\
\kappa C(\cosh\kappa A+\sinh\kappa A)+\kappa D(\sinh\kappa A+\cosh\kappa A) & =(\kappa+ik)te^{ikA},\nonumber \\
\kappa(C+D)e^{\kappa A} & =(\kappa+ik)te^{ikA}.
\end{align}
Subtracting similarly gives $\kappa(C-D)e^{-\kappa A}=(\kappa-ik)te^{ikA}$.
Solving for $C$ and $D$:
\begin{equation}
C=\frac{te^{ikA}}{2\kappa}\left[(\kappa+ik)e^{-\kappa A}+(\kappa-ik)e^{\kappa A}\right],\quad D=\frac{te^{ikA}}{2\kappa}\left[(\kappa+ik)e^{-\kappa A}-(\kappa-ik)e^{\kappa A}\right].
\end{equation}
Substituting these into the relation $2ik=ikC+\kappa D$ and simplifying
leads to the inverse transmission amplitude:
\begin{equation}
\frac{1}{t}=e^{ikA}\left[\cosh(\kappa A)+i\frac{k^{2}-\kappa^{2}}{2k\kappa}\sinh(\kappa A)\right].
\end{equation}
The transmission coefficient $T=|t|^{2}$ is then:
\begin{equation}
T=\left|\cosh(\kappa A)+i\frac{k^{2}-\kappa^{2}}{2k\kappa}\sinh(\kappa A)\right|^{-2}=\left[\cosh^{2}(\kappa A)+\left(\frac{k^{2}-\kappa^{2}}{2k\kappa}\right)^{2}\sinh^{2}(\kappa A)\right]^{-1}.
\end{equation}
\end{widetext}

\begin{figure}
\begin{centering}
\includegraphics[scale=0.3]{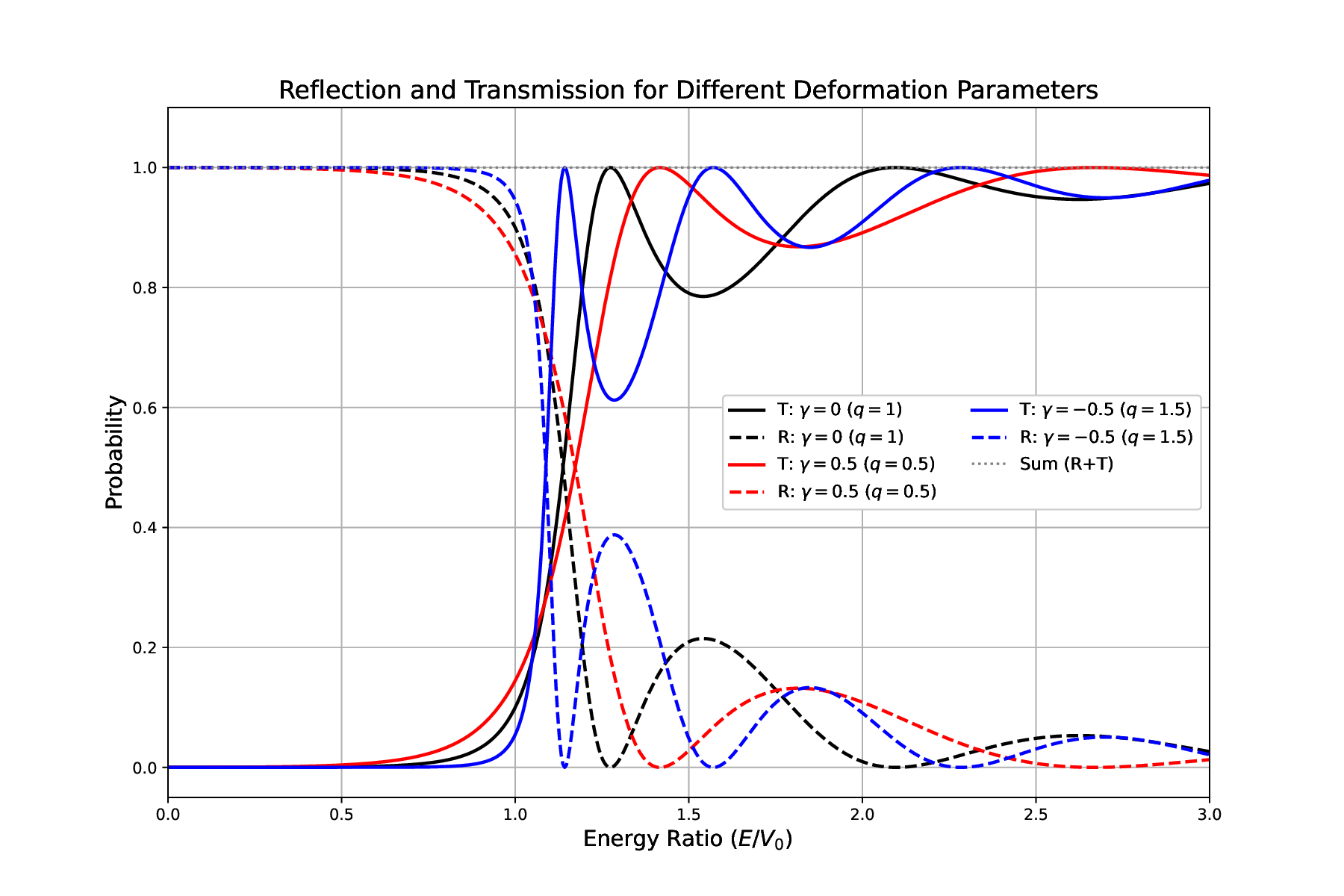}
\par\end{centering}
\caption{$q$-Modulated Tunneling Probability. Transmission coefficient
$T$ versus energy ratio $E/V_{0}$ for a barrier opacity $a\sqrt{2mV_{0}}/\hbar=6$.
For $q=0.5$ (red dotted), the effective barrier width $A(q)$ is
reduced, leading to enhanced tunneling and broader resonance peaks.
For $q=1.5$ (blue dashed), the effective width is increased, resulting
in suppressed tunneling and sharper resonances compared to the standard
$q=1.0$ case (black solid)}\label{fig:tunneling}
\end{figure}

Using $\cosh^{2}x=1+\sinh^{2}x$ and algebraic simplification of the
coefficient of $\sinh^{2}(\kappa A)$:
\begin{equation}
1+\left[1+\left(\frac{k^{2}-\kappa^{2}}{2k\kappa}\right)^{2}\right]\sinh^{2}(\kappa A)=1+\frac{(k^{2}+\kappa^{2})^{2}}{4k^{2}\kappa^{2}}\sinh^{2}(\kappa A).
\end{equation}
Substituting back the physical parameters, we arrive at the exact
expression for tunneling. For incident particles with energy $E<V_{0}$,
the transmission coefficient is
\begin{equation}
T(E;q)=\left[1+\frac{V_{0}^{2}}{4E\,(V_{0}-E)}\,\sinh^{2}\!\big(\kappa\,A(q)\big)\right]^{-1},\label{eq:Tunneling-q}
\end{equation}
with
\begin{equation}
\kappa=\frac{\sqrt{2m(V_{0}-E)}}{\hbar}.
\end{equation}
The transmission probability is plotted in Fig.~\ref{fig:tunneling}.
The deformation parameter $q$ effectively acts as a tuner for the
barrier opacity. For $q<1$, $A(q)<a$, and the barrier appears thinner
to the particle, significantly enhancing the tunneling probability
(analogous to $\gamma>0$ in Ref.~\cite{CostaFilho2011}).
For $q>1$, $A(q)>a$, and the barrier is effectively wider, suppressing
tunneling. This dependence on the effective width $A(q)$ is directly
analogous to the role played by the deformed length in the Costa Filho
formalism~\cite{CostaFilho2011}. The physical picture is transparent:
\begin{itemize}
\item For $q<1$ one has $A(q)<a$, so the barrier is effectively \emph{compressed}.
The argument of the hyperbolic sine is smaller, and the transmission
coefficient is enhanced relative to the undeformed case. This corresponds
to the behavior for $\gamma>0$ in Ref.~\cite{CostaFilho2011}.
\item For $q>1$ one has $A(q)>a$, and the barrier is effectively \emph{wider}.
The tunneling probability is accordingly suppressed, corresponding
to $\gamma<0$.
\end{itemize}
For $E>V_{0}$, the hyperbolic functions in Eq.~\eqref{eq:Tunneling-q}
are replaced by trigonometric functions with a real wave number inside
the barrier, and resonant transmission peaks shift as $A(q)$ varies,
again in full analogy with the position-dependent mass model generated
by nonadditive translations.

While the transmission coefficient $T$ characterizes the tunneling
probability, the reflection coefficient $R$ provides complementary
insight into the scattering process. Using the boundary conditions
at $\xi=0$ and $\xi=A(q)$, we can explicitly solve for the reflection
amplitude $r$. The resulting probability $R=|r|^{2}$ is given by:
\begin{equation}
R(E;q)=\frac{\frac{V_{0}^{2}}{4E(V_{0}-E)}\sinh^{2}\big(\kappa A(q)\big)}{1+\frac{V_{0}^{2}}{4E(V_{0}-E)}\sinh^{2}\big(\kappa A(q)\big)}.\label{eq:Reflection_q}
\end{equation}
This expression highlights that the reflection probability is determined
entirely by the effective barrier width $A(q)$.

The physical interpretation of $R$ in the dual-$q$ formalism is
straightforward:
\begin{itemize}
\item Suppressed Reflection ($q<1$): When the deformation parameter is less
than unity, the spatial coordinate is effectively compressed ($A(q)<a$).
The barrier appears ``thinner'' to the incident
particle, reducing the likelihood of reflection and enhancing transmission.
\item Enhanced Reflection ($q>1$): Conversely, for $q>1$, the coordinate
dilation increases the effective barrier width ($A(q)>a$), thereby
increasing the reflection probability.
\end{itemize}
Furthermore, summing the expressions for reflection (Eq.~\eqref{eq:Reflection_q})
and transmission (Eq.~\eqref{eq:Tunneling-q})
confirms that the unitarity condition is strictly preserved in the
deformed space:
\begin{equation}
R(E;q)+T(E;q)=1.
\end{equation}
This conservation of probability serves as a consistency check for
our linearization map, confirming that the dual-$q$ derivative formalism
constitutes a unitary quantum theory when the metric factors are correctly
incorporated into the definition of the physical momentum.

\subsection{Dual-$q$ harmonic oscillator in the weak-deformation regime ($q\simeq 1$)}
\subsubsection{Linearization of the Dual $q$-Schr\"odinger Equation}
The \textbf{dual $q$-derivative} is defined by
\begin{equation}
    D^{(q)} f(x) \equiv \frac{1}{1+(1-q)f(x)} \frac{df}{dx},
\end{equation}
which is inherently nonlinear in the function $f(x)$. To recover a \textit{linear} Schr\"odinger problem, we introduce the following transformations:

\paragraph{Deformed Coordinate.}
We define a new coordinate $\xi(x)$, which is invertible in the domain where $1+(1-q)x > 0$:
\begin{equation}
    \xi(x) = \frac{1}{1-q}\ln\big[1+(1-q)x\big],
    \qquad \implies \qquad
    \frac{d}{dx} = \frac{1}{1+(1-q)x}\frac{d}{d\xi}.
\end{equation}

\paragraph{Field Transformation.}
We introduce an auxiliary wave function $\chi(x)$ defined by:
\begin{equation}
    \chi(x) = \frac{1}{1-q}\ln\big[1+(1-q)\psi(x)\big],
    \qquad
    \psi(x) = \frac{e^{(1-q)\chi(x)}-1}{1-q}.
\end{equation}
This mapping is specifically chosen such that the action of the dual derivative $D^{(q)}$ on $\psi$ transforms into an ordinary derivative of $\chi$.

Consequently, the time-independent problem transforms into an \textbf{ordinary linear Schr\"odinger equation} in the space $\xi$:
\begin{equation}
    \left[-\frac{\hbar^2}{2m}\frac{d^2}{d\xi^2} + V\big(x(\xi)\big)\right]\chi(\xi) = E\,\chi(\xi),
\end{equation}
where the inverse coordinate map is given by:
\begin{equation}
    x(\xi) = \frac{e^{(1-q)\xi}-1}{1-q}.
\end{equation}

If we retain the standard physical definition of the harmonic oscillator potential,
$V(x)=\frac{1}{2}m\omega^2x^2$, the effective potential in the deformed coordinate is
\begin{equation}
    V(\xi)=\frac{1}{2}m\omega^2
    \left(\frac{e^{(1-q)\xi}-1}{1-q}\right)^2.
\end{equation}
This potential is not exactly quadratic for $q\neq1$. Moreover, it tends to a
finite plateau on one side of the $\xi$ axis and grows exponentially on the
other side. Therefore, the full problem is not the standard harmonic
oscillator and its spectrum cannot be assumed to be an infinite equally spaced
discrete sequence. The oscillator formulas below are consequently a
weak-deformation, low-lying-level approximation. They are valid only when the
states that contribute significantly satisfy the perturbative condition stated
below.
Now, we set
\begin{equation}
\gamma \equiv 1-q, \qquad |\gamma|\ll 1,
\end{equation}
and use the dual-$q$ coordinate and field maps
\begin{align}
\xi(x) &= \frac{1}{\gamma}\ln\!\big(1+\gamma x\big),
\qquad
x(\xi)=\frac{e^{\gamma \xi}-1}{\gamma},
\label{eq:xi_map}
\\[4pt]
\chi(x) &= \frac{1}{\gamma}\ln\!\big(1+\gamma \psi(x)\big),
\qquad
\psi(x)=\frac{e^{\gamma \chi(x)}-1}{\gamma}.
\label{eq:chi_map}
\end{align}
The auxiliary function $\chi(\xi)$ obeys an ordinary (linear) Schr\"odinger equation
\begin{equation}
\left[
-\frac{\hbar^2}{2m}\frac{d^2}{d\xi^2}+V\!\big(x(\xi)\big)
\right]\chi(\xi)=E\,\chi(\xi).
\label{eq:linear_Schr_xi}
\end{equation}

\subsubsection{Physical harmonic oscillator potential and small-$\gamma$ expansion}

Take the physical harmonic oscillator potential
\begin{equation}
V(x)=\frac{1}{2}m\omega^2 x^2.
\label{eq:Vx_HO}
\end{equation}
Expanding \eqref{eq:xi_map} for $|\gamma|\ll 1$ gives
\begin{equation}
\xi = x-\frac{\gamma}{2}x^2+\frac{\gamma^2}{3}x^3+\mathcal{O}(\gamma^3),
\qquad
x = \xi+\frac{\gamma}{2}\xi^2+\frac{\gamma^2}{6}\xi^3+\mathcal{O}(\gamma^3).
\label{eq:x_xi_expand}
\end{equation}
Note that, in order to obtain the energy correctly to $\mathcal{O}(\gamma^2)$, the inverse map $x(\xi)$ must be retained to one order higher than $\xi(x)$, because $V\propto x^2$ involves the square of this expansion and the cross term between $\xi$ and $(\gamma^2/6)\xi^3$ produces an additional $\mathcal{O}(\gamma^2)$ contribution. Squaring the expansion of $x(\xi)$ and collecting terms through $\mathcal{O}(\gamma^2)$ yields
\begin{equation}
x(\xi)^2 = \xi^2+\gamma\,\xi^3+\frac{7}{12}\gamma^2\,\xi^4+\mathcal{O}(\gamma^3),
\label{eq:x2_expand}
\end{equation}
where the coefficient $7/12=1/3+1/4$ comes from the cross product $2\xi\cdot(\gamma^2/6)\xi^3$ together with the square $\bigl[(\gamma/2)\xi^2\bigr]^2$. Hence
\begin{equation}
V\!\big(x(\xi)\big)
=\frac{1}{2}m\omega^2\xi^2
+\gamma\,\frac{1}{2}m\omega^2\xi^3
+\gamma^2\,\frac{7}{24}m\omega^2\xi^4
+\mathcal{O}(\gamma^3).
\label{eq:Vxi_expand}
\end{equation}
Therefore, the Hamiltonian in $\xi$-space reads
\begin{equation}
H = H_0 + \gamma H_1 + \gamma^2 H_2 + \mathcal{O}(\gamma^3),
\label{eq:H0H1}
\end{equation}
where
\begin{equation}
H_0=\frac{p_\xi^2}{2m}+\frac{1}{2}m\omega^2\xi^2,
\qquad
H_1=\frac{1}{2}m\omega^2\xi^3,
\qquad
H_2=\frac{7}{24}m\omega^2\xi^4.
\end{equation}

\subsubsection{Energy spectrum up to $\mathcal{O}(\gamma^2)$}

The unperturbed spectrum is the standard one,
\begin{equation}
E_n^{(0)}=\hbar\omega\left(n+\frac{1}{2}\right),
\qquad n=0,1,2,\dots
\label{eq:En0}
\end{equation}
The first-order shift from $\gamma H_1$ vanishes by parity,
\begin{equation}
\Delta E_n^{(1,H_1)}=\gamma\langle n|H_1|n\rangle=0,
\label{eq:En1_zero}
\end{equation}
since $H_1\propto \xi^3$ is odd. According to Rayleigh--Schr\"odinger perturbation theory, the full $\mathcal{O}(\gamma^2)$ correction receives \emph{two} contributions: the first-order shift produced by $\gamma^2 H_2$ and the second-order shift produced by $\gamma H_1$,
\begin{equation}
\Delta E_n^{(2)}=\gamma^2\,\langle n|H_2|n\rangle+\gamma^2\sum_{m\neq n}\frac{|\langle m|H_1|n\rangle|^{2}}{E_n^{(0)}-E_m^{(0)}}.
\label{eq:En2_RS}
\end{equation}
Using the standard harmonic-oscillator matrix elements $\langle n|\xi^4|n\rangle=\big(\hbar/(2m\omega)\big)^{2}(6n^{2}+6n+3)$, the first-order contribution from $H_2$ reads
\begin{equation}
\gamma^2\,\langle n|H_2|n\rangle=\gamma^2\,\frac{7}{24}m\omega^2\,\langle n|\xi^4|n\rangle=+\gamma^2\,\frac{\hbar^2}{32m}\big(14n^2+14n+7\big).
\label{eq:H2_contrib}
\end{equation}
The second-order contribution generated by $H_1$ couples only the states $n\to n\pm 1,\,n\pm 3$ and evaluates to
\begin{equation}
\gamma^2\sum_{m\neq n}\frac{|\langle m|H_1|n\rangle|^{2}}{E_n^{(0)}-E_m^{(0)}}=-\gamma^2\,\frac{\hbar^2}{32m}\big(30n^2+30n+11\big).
\label{eq:H1_secondorder}
\end{equation}
Summing the two contributions gives the corrected leading-order shift,
\begin{equation}
E_n \;=\; E_n^{(0)}+\Delta E_n^{(2)}+\mathcal{O}(\gamma^3),
\end{equation}
with
\begin{equation}
\Delta E_n^{(2)}=
-\gamma^2\,\frac{\hbar^2}{32m}\Big(16n^2+16n+4\Big)
=-\gamma^2\,\frac{\hbar^2}{8m}\,(2n+1)^{2}.
\label{eq:En2}
\end{equation}
The closed-form expression on the right exhibits a compact $(2n+1)^{2}$ dependence on the quantum number, a feature that is hidden when only the second-order $H_1$ shift is retained. The omission of $H_2$ would lead to a sign-preserving but quantitatively incorrect coefficient $-(30n^2+30n+11)$; the consistent computation requires both contributions in Eq.~\eqref{eq:En2_RS}.

\subsubsection{Auxiliary eigenfunctions up to $\mathcal{O}(\gamma)$}

Let $\phi_n(\xi)$ be the normalized eigenfunctions of $H_0$ (Hermite--Gaussian),
\begin{equation}
\begin{aligned}
\phi_n(\xi)
&=\left(\frac{\alpha}{\pi}\right)^{1/4}\frac{1}{\sqrt{2^n n!}}\,
e^{-\alpha\xi^2/2}\,
H_n\!\big(\sqrt{\alpha}\,\xi\big),
\\
\alpha
&=\frac{m\omega}{\hbar}.
\end{aligned}
\label{eq:phi_n}
\end{equation}
Then the first-order corrected auxiliary state is
\begin{equation}
\chi_n(\xi)\simeq \phi_n(\xi)
+\gamma\sum_{m\neq n}\frac{\langle m|H_1|n\rangle}{E_n^{(0)}-E_m^{(0)}}\,\phi_m(\xi).
\label{eq:chi_pert_general}
\end{equation}
Because $\xi^3$ connects only $n\to n\pm 1,n\pm 3$, one can write explicitly:
\begin{equation}
\begin{aligned}
\chi_n(\xi) &\simeq \phi_n(\xi)
+\gamma\,\frac{\sqrt{\hbar/(m\omega)}}{2^{5/2}}
\bigg[
3n\sqrt{n}\,\phi_{n-1}(\xi)
\\
& \quad -3(n+1)\sqrt{n+1}\,\phi_{n+1}(\xi)
\\
& \quad +\frac{1}{3}\sqrt{n(n-1)(n-2)}\,\phi_{n-3}(\xi)
\\
& \quad -\frac{1}{3}\sqrt{(n+1)(n+2)(n+3)}\,\phi_{n+3}(\xi)
\bigg]
\end{aligned}
\label{eq:chi_n_explicit}
\end{equation}
with the convention $\phi_{k<0}\equiv 0$.

\subsubsection{Transformation back to the physical wave function $\psi_n(x)$}

Use $\xi=\xi(x)$ from \eqref{eq:xi_map} (or its expansion \eqref{eq:x_xi_expand}) to write
\begin{equation}
\chi_n(x)=\chi_n\!\big(\xi(x)\big).
\end{equation}
From the inverse field map \eqref{eq:chi_map}, for $|\gamma|\ll 1$,
\begin{equation}
\psi(x)=\frac{e^{\gamma\chi(x)}-1}{\gamma}
=\chi(x)+\frac{\gamma}{2}\chi(x)^2+\mathcal{O}(\gamma^2),
\end{equation}
so that
\begin{equation}
\begin{aligned}
\psi_n(x)\simeq {}&
\chi_n\!\big(\xi(x)\big)
+\frac{\gamma}{2}\,\chi_n^{2}\!\big(\xi(x)\big)
\\
&+\mathcal{O}(\gamma^2).
\end{aligned}
\label{eq:psi_n_final}
\end{equation}

\paragraph{Normalization measure.}
Since $d\xi = dx/(1+\gamma x)$ from \eqref{eq:xi_map}, the natural inner product inherited from $\xi$
corresponds to the weighted measure $dx/(1+\gamma x)$ in $x$-space.

\subsubsection{Thermal properties}

In the weak-deformation regime, the low-lying bound-state spectrum of the
physical harmonic oscillator is
\begin{equation}
E_n^{(q)} \simeq \hbar\omega\left(n+\frac{1}{2}\right)
-(1-q)^2\frac{\hbar^2}{8m}(2n+1)^2,
\qquad n=0,1,2,\ldots .
\label{eq:HO_spectrum_thermal}
\end{equation}
This expression is not an exact global spectrum. It is the Rayleigh--Schr\"odinger
result obtained from the expansion of $V(x(\xi))$ through order
$\gamma^2=(1-q)^2$. The correction is negative and quadratic in
$\gamma$, so the leading shift is independent of the sign of $1-q$.
Its domain of validity is restricted by
\begin{equation}
(1-q)^2\frac{\hbar}{m\omega}(n+1)\ll1,
\label{eq:HO_validity_n}
\end{equation}
for all levels that are appreciably populated. At the thermal level this
requires, approximately,
\begin{equation}
(1-q)^2\frac{k_BT}{m\omega^2}\ll1,
\label{eq:HO_validity_T}
\end{equation}
up to numerical factors of order unity.

Within this restricted regime, the canonical partition function may be treated
perturbatively rather than by inserting Eq.~\eqref{eq:HO_spectrum_thermal} as
an exact spectrum for all $n$. Writing $\gamma=1-q$ and
$r=e^{-\beta\hbar\omega}$, the undeformed partition function is
\begin{equation}
Z_0(\beta)=\sum_{n=0}^{\infty}e^{-\beta\hbar\omega(n+1/2)}
=\frac{e^{-\beta\hbar\omega/2}}{1-e^{-\beta\hbar\omega}}.
\end{equation}
Keeping the leading deformation correction gives
\begin{equation}
Z_q(\beta)=Z_0(\beta)\left[1+\beta\frac{\gamma^2\hbar^2}{8m}
\left\langle(2n+1)^2\right\rangle_0\right]
+\mathcal{O}(\gamma^3),
\label{eq:Zq_HO_pert}
\end{equation}
where $\langle\cdots\rangle_0$ denotes the thermal average with the
undeformed oscillator weights. The required average is
\begin{equation}
\left\langle(2n+1)^2\right\rangle_0=\frac{1+6r+r^2}{(1-r)^2}.
\end{equation}
The internal energy and specific heat are obtained in the usual way,
\begin{equation}
U_q=-\frac{\partial}{\partial\beta}\ln Z_q,
\qquad
C_{V,q}=\left(\frac{\partial U_q}{\partial T}\right)_V.
\end{equation}
For $q=1$ one recovers the standard harmonic-oscillator result
\begin{equation}
C_{V,0}=k_B\frac{(\beta\hbar\omega)^2e^{\beta\hbar\omega}}
{(e^{\beta\hbar\omega}-1)^2}.
\end{equation}
Because the correction in Eq.~\eqref{eq:HO_spectrum_thermal} lowers the
low-lying energy levels, the deformation enhances the thermal accessibility
of excited states within the perturbative domain. However, the truncated
quadratic correction behaves as $-\gamma^2 n^2$ at large $n$ and is therefore
unbounded from below if extrapolated outside its validity range. For this
reason, Eq.~\eqref{eq:Zq_HO_pert} must be read as a low-temperature,
weak-deformation expansion. A full high-temperature thermodynamics would
require the exact deformed potential, including its continuum sector or an
additional box regularization.

\section{Conclusion}

We have formulated a linear quantum-mechanical treatment of the Borges dual
$q$-derivative by combining a nonlinear field transformation with the deformed
coordinate
\begin{equation}
\xi(x)=\frac{1}{1-q}\ln[1+(1-q)x].
\end{equation}
A central point of the corrected formulation is that the physical Hamiltonian
is not the formal operator $-\partial_xD^{(q)}$. Instead, the deformed
coordinate $\xi$ is taken as the canonical coordinate, with
$\hat p_\xi=-i\hbar\partial_\xi$. This choice leads to a standard linear
Schr\"odinger equation in $\xi$ and, after transforming back to the physical
coordinate $x$, to a Hermitian position-dependent-mass Hamiltonian with
\begin{equation}
 m_{\mathrm{eff}}(x)=\frac{m}{[1+(1-q)x]^2}.
\end{equation}
The associated von Roos ordering is fixed by the Jacobian of the coordinate
map and corresponds to $\alpha=\gamma=-1/4$ and $\beta=-1/2$.

For the infinite square well, the deformation replaces the physical length
$L$ by the effective length
\begin{equation}
\Xi(q)=\frac{1}{1-q}\ln[1+(1-q)L].
\end{equation}
The bound-state spectrum is therefore
$E_n^{(q)}=\hbar^2\pi^2n^2/[2m\Xi^2(q)]$. For $q<1$ the effective interval is
compressed, $\Xi(q)<L$, and the spectrum is shifted upward. For $q>1$ the
interval is dilated, $\Xi(q)>L$, and the levels move downward. The same
geometric mechanism controls the rectangular barrier: its physical width $a$
is replaced by $A(q)=\xi(a)-\xi(0)$, so tunneling is enhanced when
$A(q)<a$ and suppressed when $A(q)>a$.

The harmonic oscillator requires more care. If the physical potential is
$V(x)=m\omega^2x^2/2$, then in the canonical coordinate it becomes
\begin{equation}
V(\xi)=\frac{1}{2}m\omega^2
\left(\frac{e^{(1-q)\xi}-1}{1-q}\right)^2,
\end{equation}
which is not a quadratic oscillator for $q\neq1$. The spectrum derived in this
work is, therefore, a weak-deformation result for the low-lying bound states.
To order $(1-q)^2$, the consistent perturbative correction is
\begin{equation}
\Delta E_n^{(2)}=-(1-q)^2\frac{\hbar^2}{8m}(2n+1)^2,
\end{equation}
obtained by combining the first-order contribution of the quartic perturbation
with the second-order contribution of the cubic perturbation. The associated
thermal formulas are valid only as low-temperature perturbative expansions and
must not be extrapolated to arbitrarily high levels.

The resulting theory is isomorphic, at the geometric level, to the
nonadditive-translation formalism of Costa Filho \emph{et al.} through the
identification $\gamma=1-q$. The dual-$q$ construction therefore provides an
alternative route to the same effective geometry and PDM structure, while also
clarifying the role played by the nonlinear field map. Future work may extend
this approach to asymmetric wells, multi-barrier structures, explicitly
time-dependent systems, and many-body models in deformed geometries.

\appendix

\section{Useful properties of the $q$- and dual-$q$ derivatives}

In this appendix we summarize some algebraic and differential properties
of the $q$-calculus introduced by Borges~\cite{Borges2004}, specialized
to the notation used in the main text.

\subsection{$q$-logarithm and $q$-exponential}

The $q$-logarithm and $q$-exponential are defined for $x>0$ by
\begin{equation}
\ln_{q}x\equiv\frac{x^{1-q}-1}{1-q},\qquad e_{q}(x)\equiv\big[1+(1-q)x\big]_{+}^{1/(1-q)},
\end{equation}
where $[A]_{+}\equiv\max\{A,0\}$. They reduce to the ordinary functions
in the limit $q\to1$:
\begin{equation}
\lim_{q\to1}\ln_{q}x=\ln x,\qquad\lim_{q\to1}e_{q}(x)=e^{x}.
\end{equation}
They satisfy the ``nonextensive'' composition rules
\begin{equation}
\ln_{q}(xy)=\ln_{q}x+\ln_{q}y+(1-q)\,\ln_{q}x\,\ln_{q}y,
\end{equation}
\begin{equation}
e_{q}(x)\,e_{q}(y)=e_{q}\big(x\oplus_{q}y\big),
\end{equation}
where $\oplus_{q}$ is the $q$-sum defined below.

\subsection{$q$-sum and $q$-product}

The $q$-sum between real numbers $x$ and $y$ is defined as
\begin{equation}
x\oplus_{q}y\equiv x+y+(1-q)\,xy.
\end{equation}
It is commutative and associative, with neutral element $0$:
\begin{equation}
x\oplus_{q}y=y\oplus_{q}x,\qquad x\oplus_{q}0=x.
\end{equation}
A $q$-product is defined for $x>0$, $y>0$ by
\begin{equation}
x\otimes_{q}y\equiv\big[x^{1-q}+y^{1-q}-1\big]_{+}^{1/(1-q)}.
\end{equation}
It is also commutative and associative, with neutral element $1$:
\begin{equation}
x\otimes_{q}y=y\otimes_{q}x,\qquad x\otimes_{q}1=x.
\end{equation}
These operations allow compact expressions such as
\begin{equation}
\ln_{q}(xy)=\ln_{q}x\oplus_{q}\ln_{q}y,\qquad e_{q}(x)\,e_{q}(y)=e_{q}(x\oplus_{q}y),
\end{equation}
\begin{equation}
\ln_{q}(x\otimes_{q}y)=\ln_{q}x+\ln_{q}y,\qquad e_{q}(x)\otimes_{q}e_{q}(y)=e_{q}(x+y),
\end{equation}
valid in the domains where the deformed operations are defined.

\subsection{Standard $q$-derivative $D_{(q)}$ and $q$-integral}

The \emph{standard} $q$-derivative (used in the main text as $D_{(q)}$)
is defined by
\begin{equation}
D_{(q)}f(x)\equiv\big[1+(1-q)x\big]\,\frac{df(x)}{dx}.\label{eq:app-Dq}
\end{equation}
It satisfies
\begin{equation}
\lim_{q\to1}D_{(q)}f(x)=\frac{df}{dx},
\end{equation}
and the $q$-exponential is an eigenfunction in the unscaled case:
\begin{equation}
D_{(q)}\,e_q(x)=e_q(x).
\end{equation}
For a scaled argument $e_q(\lambda x)$, the eigenvalue relation is not
preserved by the operator $D_{(q)}=[1+(1-q)x]d/dx$ unless the deformation of
the argument is rescaled consistently. The corresponding $q$-integral
is defined (up to an additive constant) by
\begin{equation}
\int_{(q)}f(x)\,d_{q}x\equiv\int\frac{f(x)}{1+(1-q)x}\,dx,
\end{equation}
with
\begin{equation}
d_{q}x\equiv\frac{dx}{1+(1-q)x}.
\end{equation}
These operations are mutually inverse:
\begin{equation}
\int_{(q)}D_{(q)}f(x)\,d_{q}x=f(x)+\text{const}.
\end{equation}
In particular,
\begin{equation}
\frac{d}{dx}\,\ln_q x=x^{-q},
\qquad
D_{(q)}\big[\ln_q x\big]=[1+(1-q)x]x^{-q}.
\end{equation}
The standard $q$-derivative obeys the usual product rule
\begin{equation}
D_{(q)}[f(x)g(x)]=D_{(q)}f(x)\,g(x)+f(x)\,D_{(q)}g(x).
\end{equation}

\subsection{Dual $q$-derivative $D^{(q)}$ and inverse map}

The \emph{dual} $q$-derivative, denoted $D^{(q)}$ in the main text, is
\begin{equation}
D^{(q)}f(x)\equiv\frac{1}{1+(1-q)f(x)}\frac{df(x)}{dx}.
\label{eq:app-Ddual}
\end{equation}
It reduces to the ordinary derivative when $q\to1$:
\begin{equation}
\lim_{q\to1}D^{(q)}f(x)=\frac{df}{dx}.
\end{equation}
Unlike $D_{(q)}$, this operator is nonlinear in the function on which it
acts. Its inverse is therefore not an ordinary linear integral. If
\begin{equation}
D^{(q)}f(x)=g(x),
\end{equation}
then
\begin{equation}
\frac{d}{dx}\left[\frac{1}{1-q}\ln\big(1+(1-q)f(x)\big)\right]=g(x),
\end{equation}
and hence
\begin{equation}
f(x)=\frac{1}{1-q}\left\{C\exp\!\left[(1-q)\int^x g(s)\,ds\right]-1\right\},
\end{equation}
where $C$ is fixed by the boundary or initial condition. This is the origin
of the field map used in Eq.~\eqref{eq:chiDef}.

The standard and dual derivatives are related by
\begin{equation}
D^{(q)}f(x)=\frac{1}{[1+(1-q)x][1+(1-q)f(x)]}\,D_{(q)}f(x),
\label{eq:app-Drelation}
\end{equation}
which expresses the dual nature of the two operators. The product rule for
the dual derivative is
\begin{align}
D^{(q)}[f(x)g(x)] &=\frac{1}{1+(1-q)f(x)g(x)}
\Big\{[1+(1-q)f(x)]D^{(q)}f(x)\,g(x)
\nonumber\\
&\qquad\qquad\qquad+[1+(1-q)g(x)]f(x)\,D^{(q)}g(x)\Big\}.
\end{align}

\subsection{Chain rule and composition}

For a composition $F\!\circ G$ one may write, whenever the functions
are sufficiently regular,
\begin{equation}
D_{(q)}[F(G(x))]=\big[1+(1-q)x\big]\,F'(G(x))\,G'(x),
\end{equation}
\begin{equation}
D^{(q)}[F(G(x))]=\frac{F'(G(x))\,G'(x)}{1+(1-q)\,F(G(x))},
\end{equation}
with the obvious $q\to1$ limits. These relations provide the basic
toolbox for handling $q$-exponentials, $q$-logarithms and the associated
deformed derivatives used throughout the main text.

\section{Hermiticity of the canonical dual-$q$ Hamiltonian}

In the main text the physical Hamiltonian is defined in the canonical
coordinate $\xi$ as
\begin{equation}
\hat H_\xi\chi(\xi)=\left[-\frac{\hbar^{2}}{2m}\frac{d^{2}}{d\xi^{2}}
+V\big(x(\xi)\big)\right]\chi(\xi)=E\chi(\xi).
\label{eq:app-Hxi}
\end{equation}
This appendix proves the Hermiticity of this Hamiltonian and clarifies its
relation to the formal operator obtained by direct substitution of
$D^{(q)}$.

\subsection{Hermiticity in the $\xi$ representation}

With the scalar product
\begin{equation}
\langle\chi_1|\chi_2\rangle_\xi=\int_{I_\xi}d\xi\,\chi_1^*(\xi)\chi_2(\xi),
\end{equation}
and boundary conditions that remove surface terms, integration by parts gives
\begin{align}
\langle\chi_1|\hat H_\xi\chi_2\rangle_\xi
&=-\frac{\hbar^2}{2m}\int_{I_\xi}d\xi\,\chi_1^*\frac{d^2\chi_2}{d\xi^2}
+\int_{I_\xi}d\xi\,\chi_1^*V\chi_2
\nonumber\\
&=\langle\hat H_\xi\chi_1|\chi_2\rangle_\xi .
\end{align}
Thus $\hat H_\xi$ is Hermitian in the canonical representation.

\subsection{Equivalent Hermitian form in the $x$ representation}

The measures are related by
\begin{equation}
d\xi=\frac{dx}{w(x)},
\qquad w(x)=1+(1-q)x.
\end{equation}
Defining $\chi(\xi(x))=w^{1/2}(x)\phi(x)$ maps the scalar product to the
flat $x$-space measure,
\begin{equation}
\int d\xi\,|\chi(\xi)|^2=\int dx\,|\phi(x)|^2.
\end{equation}
Under this unitary map the kinetic energy becomes
\begin{equation}
\hat K_x\phi(x)=
-\frac{\hbar^{2}}{2m}\,w^{1/2}(x)\frac{d}{dx}
\left[w(x)\frac{d}{dx}\left(w^{1/2}(x)\phi(x)\right)\right],
\end{equation}
which is exactly the PDM operator in Eq.~\eqref{eq:PDM_Final}. Therefore the
Hermiticity of the $x$-space Hamiltonian follows from the Hermiticity of
$\hat H_\xi$ and the unitarity of the measure transformation.

It is useful to distinguish this physical Hamiltonian from the formal direct
operator
\begin{equation}
\widetilde K^{(q)}\psi(x)=
-\frac{\hbar^{2}}{2m}\partial_x\big[D^{(q)}\psi(x)\big].
\end{equation}
Using $D^{(q)}\psi=d\chi/dx=w^{-1}d\chi/d\xi$, this operator becomes
\begin{equation}
\widetilde K^{(q)}\psi(x)=
-\frac{\hbar^{2}}{2m}\frac{1}{w(x)}\partial_\xi
\left[\frac{1}{w(x)}\partial_\xi\chi(\xi)\right],
\end{equation}
which is not the same as $-(\hbar^2/2m)\partial_\xi^2\chi$. This is why the
present paper adopts the canonical $\xi$-Hamiltonian as the physical starting
point and regards the direct $-\partial_xD^{(q)}$ expression only as a formal
motivation.

\section{Continuity equation in the dual-$q$ formalism}

We now provide the detailed derivation of the continuity equation
in the dual-$q$ framework, starting from the linearized Schr\"odinger
equation in $\xi$ and translating the result to the $x$ representation.

\subsection{Continuity equation in the $\xi$ representation}

The time-dependent Schr\"odinger equation in the deformed coordinate
is
\begin{equation}
i\hbar\,\frac{\partial\chi}{\partial t}=\hat{H}_{\xi}\chi=-\frac{\hbar^{2}}{2m}\frac{d^{2}\chi}{d\xi^{2}}+V\big(x(\xi)\big)\chi.\label{eq:app-TDSE-xi}
\end{equation}
Its complex conjugate reads
\begin{equation}
-i\hbar\,\frac{\partial\chi^{*}}{\partial t}=-\frac{\hbar^{2}}{2m}\frac{d^{2}\chi^{*}}{d\xi^{2}}+V\big(x(\xi)\big)\chi^{*}.
\end{equation}
Multiplying Eq.~\eqref{eq:app-TDSE-xi} by $\chi^{*}$, the conjugate
equation by $\chi$, and subtracting, we obtain
\begin{equation}
i\hbar\,\frac{\partial}{\partial t}|\chi|^{2}=-\frac{\hbar^{2}}{2m}\frac{d}{d\xi}\left(\chi^{*}\frac{d\chi}{d\xi}-\chi\frac{d\chi^{*}}{d\xi}\right).
\end{equation}
Defining the probability density and current in the $\xi$ representation
as
\begin{equation}
\rho_{\xi}(\xi,t)=|\chi(\xi,t)|^{2},
\end{equation}
\begin{equation}
j_{\xi}(\xi,t)=\frac{\hbar}{m}\,\Im\!\left[\chi^{*}(\xi,t)\,\frac{d\chi}{d\xi}(\xi,t)\right],
\end{equation}
we obtain the standard continuity equation
\begin{equation}
\frac{\partial\rho_{\xi}}{\partial t}+\frac{\partial j_{\xi}}{\partial\xi}=0.\label{eq:app-cont-xi}
\end{equation}

\subsection{Continuity equation in the $x$ representation}

Using the relation
\begin{equation}
\frac{d}{d\xi}=\big[1+(1-q)x\big]\frac{d}{dx}\equiv w(x)\frac{d}{dx},
\end{equation}
we can rewrite Eq.~\eqref{eq:app-cont-xi} as
\begin{equation}
\frac{\partial}{\partial t}\left(\frac{\rho_{\xi}}{w(x)}\right)
+\frac{\partial j_{\xi}}{\partial x}=0.
\end{equation}
Thus, the probability density with respect to the flat measure $dx$ is
\begin{equation}
\rho_x(x,t)=\frac{|\chi(\xi(x),t)|^{2}}{w(x)}=|\phi(x,t)|^{2},
\end{equation}
and the corresponding current is
\begin{equation}
J_x(x,t)=j_{\xi}(\xi(x),t)
=\frac{\hbar}{m}\,\Im\!\left[\chi^{*}(\xi(x),t)\frac{d\chi}{d\xi}(\xi(x),t)\right].
\end{equation}
These quantities satisfy the physical continuity equation
\begin{equation}
\frac{\partial\rho_x}{\partial t}+\frac{\partial J_x}{\partial x}=0.\label{eq:app-cont-x}
\end{equation}
Using the identity $D^{(q)}\psi=d\chi/dx$ and Eq.~\eqref{eq:dxdxi},
we can also write
\begin{equation}
\frac{d\chi}{d\xi}=\big[1+(1-q)x\big]\,\frac{d\chi}{dx}
=\big[1+(1-q)x\big]D^{(q)}\psi,
\end{equation}
so that the current can be expressed in terms of the linearized physical
field as
\begin{equation}
J_x(x,t)=\frac{\hbar}{m}\,\Im\!\left[F(\psi)^{*}\,\big[1+(1-q)x\big]D^{(q)}\psi\right].
\end{equation}
The quantity $\rho_x(x,t)$ is therefore the properly normalized
probability density in the physical coordinate, and $J_x(x,t)$ is the
associated conserved probability current.
\bibliographystyle{apsrev4-2}
\bibliography{articlebib_revised}

\end{document}